# Terahertz Spectrometer of Wavelength Dimensions Based on Extraordinary Transmission

M. Henstridge,[1,2,*] J. Zhou,[1,3] L. Jay Guo,[1,3] and R. Merlin[1,2]

[1]*Center for Photonics and Multi-Scale Nanomaterials, University of Michigan,*

*Ann Arbor, 48109, USA*

[2]*Department of Physics, University of Michigan, Ann Arbor, 48109-1040, USA*

[3]*Department of Electrical Engineering and Computer Science,*

*University of Michigan, Ann Arbor, 48109-2122, USA*

*\*Corresponding author: mhenst@umich.edu*

Subwavelength-slotted parallel plate waveguides exhibit a localized electromagnetic resonance bound to the slits at a frequency slightly below the transverse electric cutoff [R. Merlin, Phys. Rev. X **2**, 031015 (2012)]. The resonance is long-lived and, as opposed to the vanishingly small transmission shown by a single sub-wavelength aperture, it gives perfect transmission for perfectly-conducting plates. We show that the aperture-supported resonances of a pair of slotted copper plates have long lifetimes at THz frequencies. Finite element method calculations show that these bound resonances can have quality factors greater than 100. Effects of plate length and imperfect parallel alignment are also discussed. Using THz time domain spectroscopy, we measured the transmission of a broadband pulse through a test structure for several plate separations. These results suggest that the slotted waveguide can function as a highly compact THz spectrometer.

Terahertz (THz) imaging and spectroscopy provide a wealth of information to researchers from a wide range of disciplines. Filling the gap between the microwave and infrared range, THz radiation has shown to be extremely useful to probe molecular dynamics [1-3], distinguish between cancerous and benign tissue [4,5], and for the identification of contraband concealed within packages and clothing that are opaque at visible frequencies [6,7]. Efforts have also focused on the collection of extraterrestrial far-infrared and THz emissions, the analysis of which provides information regarding planetary atmospheres, the life cycle of interstellar clouds, and other phenomena of interest to astronomers and cosmologists [8-10].

One of the oldest and best-known methods for measuring THz spectra is Fourier Transform Infrared (FTIR) Spectroscopy, in which radiation is spectrally resolved with the use of, say, a Michelson interferometer [11]. A more modern technique, Terahertz Time Domain Spectroscopy (THz-TDS), is a coherent method sensitive to the electric field. In most THz-TDS setups, the emission and detection processes rely on an ultrafast laser to excite either a photoconductive antenna [12-14] nonlinear crystal [15,16] or air plasma [17,18]. Another coherent THz method involves tuning the frequency of a continuous-wave source and detecting the field transmitted or reflected from a sample [19,20]. THz applications are often limited by the relatively large size of the available systems. Here, we present a highly compact THz spectrometer which can allow for new applications, such as on-chip spectroscopy for integration with microfluidics platforms [21,22], and can easily be incorporated into any given optical setup to quickly characterize a THz source. Also, integrating a chip-sized THz spectrometer with a remotely-controlled device could allow for fast remote characterization of hazardous environments through the spectral fingerprinting of substances such as toxic gasses [23,24] and air-borne weapons [25].

In this letter, we show that a pair of parallel copper plates each containing a single sub-wavelength slit can be used as a highly-compact THz spectrometer. The principles of this structure are discussed in [26], and a microwave realization was reported in [27]. As shown for the case of perfectly conducting plates [26], such a system shows an electromagnetic resonance at a frequency slightly below the transverse electric (TE) waveguide cutoff. This resonance is supported by the slits, and unlike the extended fields of a parallel-plate Fabry-Perot spectrometer, its fields are localized within the plates. Considering that a negligible amount of radiation transmits through a single sub-wavelength-sized aperture [26,28], this aperture-bound resonance is remarkable in that it gives perfect transmission in the absence of conduction losses. Higher-order resonances supported by the slits also exist at frequencies near integers of the fundamental Fabry-Perot resonance. However, their coupling to the waveguide continua does not allow for perfect transmission. We note that these aperture-supported resonances are not excited in the transverse-magnetic (TM) - polarized configuration, for which the transmission spectrum is defined by broad peaks at the Fabry-Perot resonant frequencies [26,29].

Our slotted waveguide spectrometer consists of two polymethylpentene (TPX) substrates each containing a copper slotted plate; see Fig. 1. Samples were fabricated using a thermal evaporator to deposit a 1 μm layer of copper onto the substrates. The deposition was performed while a shadow mask, shaped into the complement of the slotted plate, was attached to each substrate. By mounting the resulting samples onto linear translation stages, we constructed a copper slotted waveguide with an adjustable air gap. An angular kinematic mount was used to achieve parallel alignment.

Finite element method calculations were carried out using the software COMSOL 3.5a. In one set of calculations, we consider an incident cylindrical Gaussian beam and solved the two-dimensional (2D) Helmholtz equation using the actual parameters of our Cu-plate structure. We also performed three-dimensional (3D) simulations assuming perfectly conducting plates to account for the finite size of, both, the plates and the illuminated area. Results of the 2D calculations are reproduced in the contour plots of Fig. 2. TE transmission data is shown in Fig. 2(a) and 2(c) for, respectively, the fundamental frequency and the first overtone. Note the field profile between the plates revealing the localized nature of the electromagnetic resonances

[26,27]. The corresponding TM plots show the transmission that results from excitation at the corresponding waveguide cut-offs.

We measured the transmission of both TE and TM polarized radiation for various plate separations using a standard THz-TDS setup. THz pulses were generated by focusing the output of a Ti:Sapphire oscillator, tuned to 815nm with a repetition rate of 82MHz, onto a biased photoconductive emitter. Using parabolic reflectors, the emitted THz pulse was focused onto the slotted plates. The radiation transmitted through the structure was spatially overlapped with a separate near-IR probe pulse from the oscillator, and then focused onto a (110) ZnTe crystal for detection via electro-optic sampling. The path of the THz radiation was sealed inside a plexi-glass enclosure filled with nitrogen gas in order to eliminate absorption by water vapor.

Figure 3(a) and 3(b) show 2D simulations at $d = 139$ µm and experimental THz-TDS traces at $d = 140$ µm for the TE and TM geometry, respectively. Experimental plate separations were obtained from the measured resonant frequencies using the Fabry-Perot condition $d = mc/2f_m$ where $f_m$ is the frequency of the resonance of order $m$ and $c$ is the speed of light in vacuum. To generate the theoretical traces, we fitted each spectrum from the simulation to a sum of Lorentzians and used the fit parameters to generate the corresponding waveforms, which were then convolved with the reference THz-TDS pulse trace. Fig. 3(c) shows the Fourier spectra for both polarizations. The theoretical spectra are from the simulations and the fits to the experimental data were obtained using a linear prediction method [30,31]. We found this method to be more accurate than a simple Fourier transformation because of the limit to the measured time window imposed by a spurious pulse in the reference trace. In Fig. 4, measured TE spectra are shown for four different plate separations. We note that the peaks have contrasts in the range $10^2$-$10^3$. The localized nature of the resonances and the evident ability to tune the peak frequencies with plate separation set our structure as a viable wavelength-size THz spectrometer. As discussed below, the quality factors ($Q$) of the resonances can be increased considerably by improving the alignment of the plates.

The comparison between theory and experiment shows that there are differences of nearly a factor of 5 between the calculations and measurements of the peak power transmission. We ascribe these differences to losses associated with the finite size of both the structure and the incident illumination profile. This assignment is supported by the 3D simulations, which consider a Gaussian beam focused onto a perfectly conducting slotted waveguide. The length of the plates along the slit and the size of the incident beam were the same as those used in the experiments. These simulations give a factor of ~ 5 (~ 25) decrease in the power transmission for the TE (TM) polarization when compared with an infinite structure. Multiplying the power spectra and field amplitudes from the calculations by the respective factors yields peak transmission values that agree reasonably well with the data, as shown in Fig. 3.

The considerable differences between the measured and theoretical quality factors in Fig. 3(c) are attributed mainly to the imperfect parallel alignment of the plates. 2D simulations at a plate separation of 167 μm show that a deviation as small as $0.1^o$ from parallel alignment decreases the quality factor of the first order resonance (Q = 850) by a factor of 3.5, whereas the second order resonance (Q =130) was not affected (however, increasing the angle to $1^o$ reduces the lifetime of the second order resonance by a factor of 2). Long-lived resonances have fields that extend farther throughout the plates compared to those of shorter-lived resonances. As such, the plate alignment is more critical for their observation.

In conclusion, we demonstrated the principles of a sub-wavelength slotted waveguide at THz frequencies by measuring both TE and TM responses of a copper test structure with THz-TDS. For the TE geometry, we tuned the frequency of high-contrast resonances with plate separation, and our simulations show that improved alignment of the plates could allow for the observation of longer-lived resonances with quality factors of almost 1000. We have shown that our structure works as a highly-compact THz spectrometer, and its further development would allow for new THz-based applications.

FUNDING

Work supported by the MRSEC Program of the National Science Foundation under Grant No. DMR-1120923.

ACKNOWLEDGMENTS

The authors would like to thank S.M. Young for advice and training with the components involved with THz-TDS.

FIGURE CAPTIONS

FIG1. Schematic diagram describing the slotted plates. The slits' width is ~ 40 μm. The plates' dimensions are 5 mm × 7 mm. The thickness of the substrates is 2 mm.

FIG2. Contour plots showing calculated TE and TM field amplitudes at a plate separation of 167μm. The frequencies are (a) $v_{TE1}$ = 0.8953 THz. (b) $v_{TM1}$ = 0.8970 THz. (c) $v_{TE2}$ = 1.778 THz. (d) $v_{TM2}$ = 1.795 THz.

FIG3. Calculated (*d* = 139 μm) and measured (*d* = 140 μm) THz-TDS traces for the TE (a) and TM (b) geometry. The corresponding power spectra are shown in (c).

FIG4. Measured TE transmission spectra for various plate separations. Arrows denote the second-order resonance.

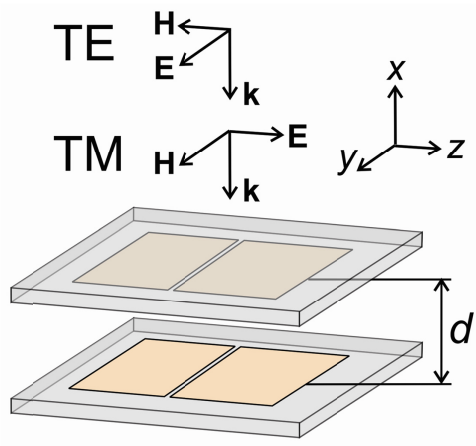

Fig 1

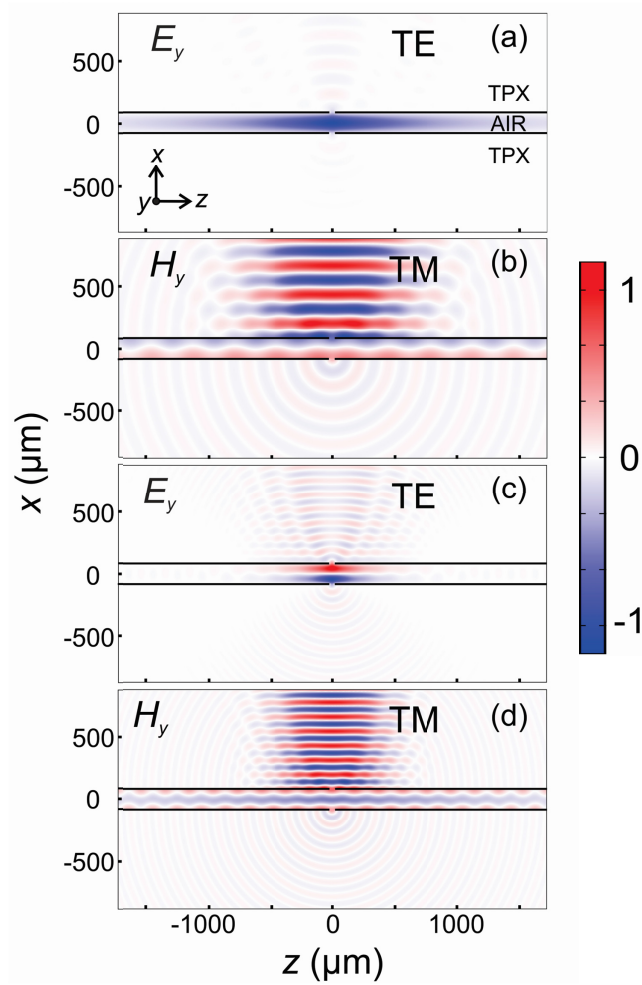

Fig 2

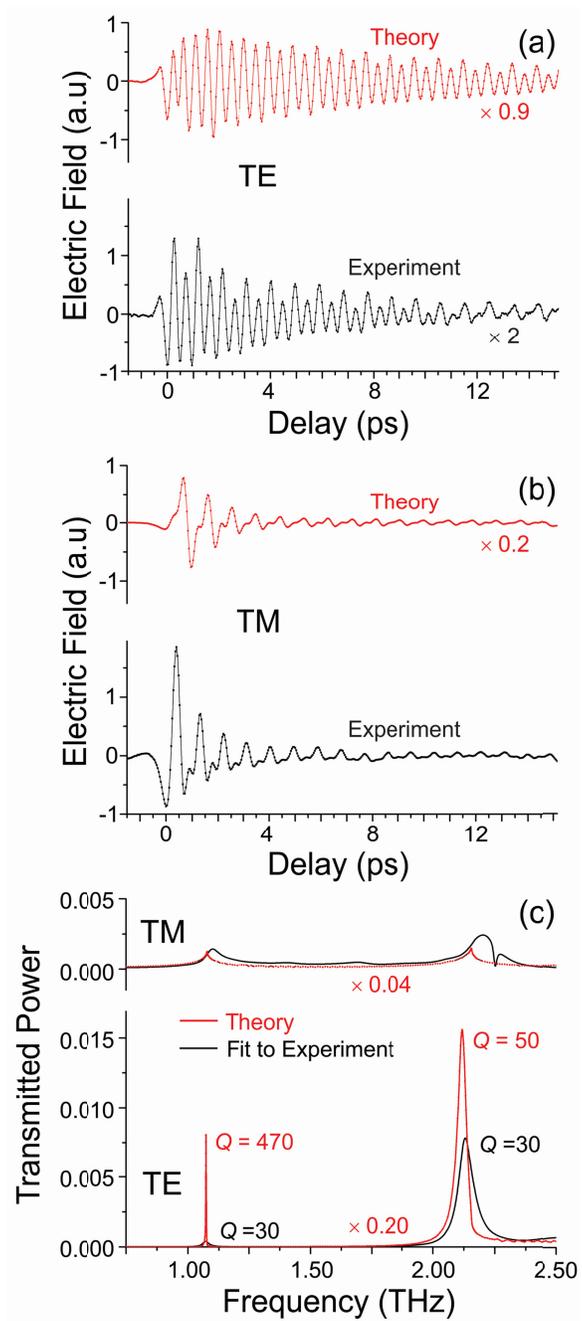

Fig 3

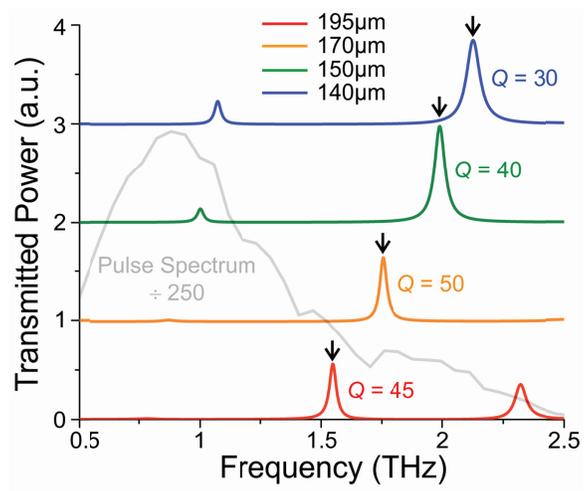

Fig 4